\documentstyle[epsf,twoside,fleqn,espcrc2,epsfig]{article}

\newcommand{\AmS}{{\protect\the\textfont2
  A\kern-.1667em\lower.5ex\hbox{M}\kern-.125emS}}

\title{Balls in Boxes and Quantum Gravity}

\author{Piotr~Bialas\address{Institute of Computer Science,
Jagellonian University, Reymonta 4, 30-059 Krak\'{o}w, Poland},
Zdzis\l{}aw~Burda\address{Institute of Physics,
Jagellonian University, Reymonta 4, 30-059 Krak\'{o}w, Poland}
and
D. A. Johnston\address{Mathematics Department, Heriot-Watt University, 
Edinburgh, EH14 4AS, United Kingdom}}

\begin{document}


\maketitle

Four dimensional simplicial gravity has been studied by means of
Monte Carlo simulations for some time  \cite{hist},
the main outcome of the studies being that the model  
undergoes a discontinuous phase transition  \cite{first}
between an elongated and a crumpled phase when one changes
the curvature (Newton) coupling.  In the
crumpled phase there are singular vertices 
growing extensively with the volume of the
system \cite{sing} giving an infinite Hausdorff dimension,
whereas 
the elongated phase has a
Hausdorff dimension equal to two. This phase has all properties of a
branched-polymer phase \cite{aj}.
We have postulated\cite{bbpt} that this behaviour is a
manifestation of the constrained-mean-field scenario as realised
in the Branched Polymer\cite{bb} (BP) or
Balls-in-Boxes model\cite{bbj}. The models of \cite{bbpt,bb,bbj}
share all the
features of 4D simplicial gravity except that they
exhibit a continuous phase transition. We note here
that this defect can be remedied by a suitable
choice of ensemble.  

The partition function of the Balls-in-Boxes model\cite{bbj} is
\begin{eqnarray}
Z(M,N)\;   = 
\sum_{\{q\}}p(q_1)\cdots p(q_M)
\delta_{q_1+\cdots+q_M,N},
\label{zmn}
\end{eqnarray}
which describes weighted partitions of $N$ balls in $M$ boxes. 
This model can be solved\cite{bbj}
in the limit of an infinite number of boxes and fixed density
of balls per box~: $M\rightarrow \infty$ and $\rho=N/M = const$. In
this limit the partition function can be expressed in terms of the
free energy density per box $f(\rho)$ 
\begin{eqnarray}
Z(M,N)= e^{M f(\rho) + \dots}.
\end{eqnarray}
By introducing the integral representation of the
Kronecker delta function one finds by the steepest descent method that
\begin{eqnarray}	
f(\rho) = \mu_*(\rho)\rho+ K(\mu_*(\rho))
\label{ffluid}
\end{eqnarray}
where $K$ is a  generating function given by
\begin{equation}	
K(\mu)=\log\sum_{q=1}^\infty p(q)e^{-\mu q}
\label{series}
\end{equation}
and $\mu_*(\rho)$ is a solution of the saddle point equation
\begin{eqnarray}	
\rho + K'(\mu_*) = 0
\label{sp}
\end{eqnarray}
For a suitable choice
of the weights $p(q)$ the system displays
a two phase structure with a critical density $\rho_{cr}$.

When $\rho$ approaches
$\rho_{cr}$ from below $\mu_*$ approaches $\mu_{cr}$ from above.  When
$\rho$ is larger than $\rho_{cr}$~, $\mu$ sticks
to the critical value $\mu_{cr}$ and the free energy is linear in
$\rho$
\begin{eqnarray}	
f(\rho) = \mu_{cr}\rho+ \kappa_{cr}
\label{fcond}
\end{eqnarray}
where $\kappa_{cr}=K(\mu_{cr})$. The change of regimes
$\rho<\rho_{cr}$ (\ref{ffluid}) to $\rho\ge\rho_{cr}$ (\ref{fcond})
corresponds to the phase transition. To understand the
physical nature of
the transition it is convenient to consider the dressed
one-box probability, defined as the
probability that a particular
box contains  $q$ balls.  
In the large $M$ limit the saddle point equation gives
\begin{equation}
\pi(q) = \left\{ \begin{array}{lll}
e^{-K(\mu_*)} p(q)e^{-\mu_* q} & \mbox{ for } & \rho < \rho_{cr} \\
&& \\
e^{-\kappa_{cr}} p(q)e^{-\mu_{cr} q} & \mbox{ for } & \rho \ge \rho_{cr}
\end{array} \right .
\label{pisp}
\end{equation}
The approach of the dressed probability to the limiting form (\ref{pisp})
is not uniform. 
In particular the average,
$\langle q \rangle = \sum_q\ q \pi(q)$, 
does not give $\rho$ for $\rho>\rho_{cr}$ as it should. One can easily
correct for this by adding a ``surplus anomaly'' term to $\pi$ for
finite $M$ in the phase above $\rho_{cr}$ \cite{bbj2}
\begin{equation}
\pi_M(q) = \pi(q) + \frac{1}{M} \delta(q - M(\rho-\rho_{cr})).
\end{equation} 
In effect, $M-1$
boxes keep the critical form of the distribution and one box takes over
the surplus of balls.  One can easily check by performing directly
finite size computations that this situation is indeed realized in the
model. 

The appearance of the surplus anomaly is a condensation
phenomenon
similar to Bose-Einstein condensation
and the transition in the Kac-Berlin spherical model\cite{Kac}. 
We call the phase $\rho<\rho_{cr}$ {\it fluid} and $\rho>\rho_{cr}$ {\it condensed}. The
system may enter the condensed phase either by changing density or by
modifying the weights $p$.  For instance, the one parameter family
of weights
$p(q) = q^{-\beta}$, $q\ge 1$
gives the critical line
\begin{equation}
\rho_{cr} = \frac{\zeta(\beta-1)}{\zeta(\beta)}
\end{equation}
so one can change the phase either by varying $\beta$ for fixed
$\rho$ (as in case of branched polymers where $\rho=2$), or by fixing
$\beta$ and varying the density $\rho$.
The transitions in $\rho$ and $\beta$ in this variant
of the model are continuous, unlike the simulations
of simplicial gravity.

\begin{figure}[t]
\begin{center}
\epsfig{file=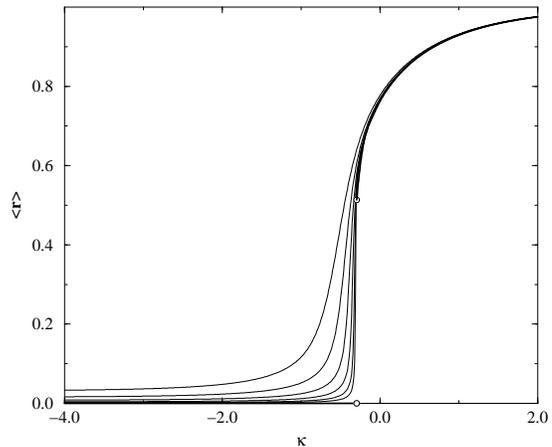,width=8cm,bbllx=18,bblly=120,bburx=552,bbury=430}
\end{center}
\caption{\label{r}The average curvature as a function of $\kappa$. The heavy
line corresponds to thermodynamical limit. ($\beta=5/2$)}
\end{figure}

It is also possible to consider ensembles with varying 
density. The simplest candidate is an  ensemble  
with a chemical potential coupled to the total number 
of {\it balls},
which can be treated as a box in contact with
a reservoir of balls.
However, this leads to a totally decoupled system
which corresponds to $M$ copies of the urn-model\cite{urn}.
For $\mu\ge\mu_{cr}$ the average number of balls in the urn diverges. 
More interesting in the context of simplicial gravity is the model where
we keep the number of balls fixed and vary the number of
{\it boxes} \cite{bbj2,bb2}. 
This gives direct analog of the 
ensemble used in the simplicial gravity 
simulations 
\begin{equation}\label{zkappa}
Z(\kappa,N) = \sum_{M} Z(M,N) e^{\kappa M}.
\label{ccan}
\end{equation}
It is now more natural to consider {\em curvature} $r=1/\rho$
instead of density. The partition function (\ref{zkappa}) 
can be rewritten as
\begin{eqnarray}\label{zkappai}
Z(\kappa,N)\approx N\int\limits_{1/N}^1\mbox{d}re^{N (f(r)+\kappa r)}
\end{eqnarray}
and the saddle point equation for this integral is
\begin{eqnarray}
\kappa + f'(r_*) = 0.
\label{fprim}
\end{eqnarray}
For $\kappa > \kappa_{cr}$ this equation
reduces to $\kappa =  K(\mu_{sp}(r_{*}))$ 
which has a unique solution for $r_*$. 
The value of $r_*$ (the centre of a gaussian distribution)
is the average curvature  
in the limit $N\rightarrow \infty$. 
This situation continues as long as $\kappa > \kappa_{cr}$.

\begin{figure}[t]
\begin{center}
\epsfig{file=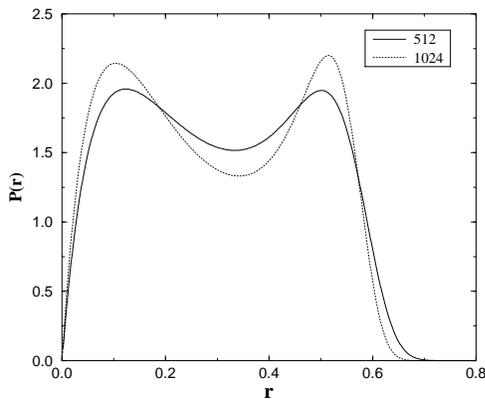,width=7cm,bbllx=18,bblly=120,bburx=552,bbury=430}
\end{center}
\caption{\label{hist}Distribution of $r$ near the phase transition for
systems with 512 ($\kappa=-0.32184$) 
and 1024 balls ($\kappa=-0.31910$) ($\beta=5/2$).
}
\end{figure}

For $\kappa < \kappa_{cr}$ the saddle point equation (\ref{fprim}) has no
solution and therefore the integrand is no longer gaussian 
but a monotonic function of $r$. In particular, for $r<r_{cr}$ it is
exponential, $\exp N(\kappa - \kappa_{cr})r$, and for large $N$ only this 
exponential part contributes in the integral (\ref{zkappai}) giving
$\langle r\rangle\sim 1/N$. 

The average curvature $\langle r \rangle$ is shown
as a function of $\kappa$ in Figure~\ref{r}.
The bold line is a limiting curve for $N=\infty$.
For $\kappa > \kappa_{cr}$ it is the solution of the saddle point 
equation (\ref{fprim}). It stops at $r_{cr}$ and falls to zero. 
In the neighbourhood of the critical point the curves are steepest.
This part of the curves corresponds to the 
pseudocritical region where the two phases coexist. One expects
a double peak histogram for $r$~: one peak near the maximum of
the gaussian phase and the other near the kinematic limit $1/N$.
In Figure~\ref{hist} we show the distributions of $r$
for two different sizes $N$, showing the coexistence of two phases
that is characteristic of first order transitions.

To summarize, the Balls-in-Boxes model 
describes well basic features of simplicial gravity 
simulations such as the appearance
of the singular vertices and the mother universe\cite{bbpt}.
With the appropriate choice of ensemble one obtains
a first order transition too.
We note that arguments \cite{ckr,ac} have recently been given that 
the bare weights $p(q)$ in simplicial gravity
can be approximated by $p(q) \sim e^{b q} e^{-a q^{\sigma}}$,
which gives a similar behaviour to the power law weights discussed
here.

This work was partially supported by the KBN grant
2P03B04412.

\end{document}